\documentclass{article}

\def\bm#1{\mbox{\boldmath{$#1$}}}   % this is used to write boldface Greek

\usepackage{amssymb}
\usepackage{amsmath}
\usepackage[dvips]{epsfig}

\newcommand{\la}{\label}

\def\rr#1{(\ref{#1})}
\renewcommand{\vec}[1]{\boldsymbol{#1}}
\newcommand{\ii}{\textrm{i}}
\newcommand{\ee}{\textrm{e}}
\newcommand{\demi}{\textstyle{\frac{1}{2}}}

\def \div{\mbox{div\hskip 1pt}}
\def \Div{\mbox{Div\hskip 1pt}}
\def \tr{\mbox{tr\hskip 1pt}}
\def \grad{\mbox{grad\hskip 1pt}}
\def \Grad{\mbox{Grad\hskip 1pt}}
\begin{document}

\title{Small amplitude waves and stability\\ for a pre-stressed
viscoelastic solid}

\author{Michel Destrade, Ray W. Ogden, Giuseppe Saccomandi}

\date{2008}

\maketitle

%%%%%%%%%%%%%%%%
\begin{abstract}

We study the propagation of small amplitude waves superimposed on
a large static deformation in a nonlinear viscoelastic material of
differential type. We use bulk waves and surface waves to address
the questions of dissipation and of material and geometric
stability. In particular, the analysis provides bounds on the
constitutive parameters and on the pre-deformation that ensure
linearized stability in the neighbourhood of a large pre-stretch.
This type of result is relevant to the imaging of biological soft
tissues using acoustical techniques, where pre-deformation is
known to increase contrast and reduce de-correlation noise.

\end{abstract}
%%%%%%%%%%%%%%

\maketitle

\section{Introduction}

In many important technological applications, polymeric materials
--- such as the elastomers used in engine mounts or bridge
bearings --- are subject to large deformations, and infinitesimal
theories are not suitable for modelling their mechanical response.
This is true also for complex biomaterials `in service' such as
ligaments, tendons, skin, arteries, and other biological soft
tissues that have several mechanical features in common with
elastomeric polymers.

An adequate modelling of rubber-like materials and of biological
soft tissues subject to large deformations requires the use of the
theories of \emph{nonlinear elasticity} and \emph{nonlinear
viscoelasticity}. Whereas the theory of nonlinear elasticity is a
well-developed chapter of solid mechanics, the theory of nonlinear
viscoelasticity is still in its infancy. Relatively few studies
have been carried out beyond establishing basic constitutive
characterizations and their general thermodynamical implications.
In particular, there is a paucity of complete studies of the
propagation of \emph{mechanical waves}. Beyond the literature
dedicated to acceleration waves and universal motions, we find few
papers dedicated to finite amplitude waves and to small amplitude
waves superimposed on finite deformations in viscoelastic solids
(of course, the situation is different for waves in viscoelastic
fluids).

Antman and Seidman \cite{Antman} provided a detailed mathematical
study of large shearing motions of nonlinearly viscoelastic slabs
(see also Rajagopal and Saccomandi \cite{Rajagopal} for some exact
solutions in a similar framework). Recently, Hayes and Saccomandi
\cite{hayes1, hayes2, hayes3} and Destrade and Saccomandi
\cite{ds1, ds2, ds3} obtained some results for finite amplitude
motions and waves in some classes of nonlinear viscoelastic
materials. Earlier, Hayes and Rivlin \cite{haye69, hay72a, hay72b,
hay72c, haye74} had established some general results for the
theory of small motions superposed on a large deformation in
nonlinear viscoelastic solids (see also a recent note by
Saccomandi \cite{sacc2002} concerning such waves in a special
class of materials).

The situation is, of course, completely different in the linear
theory of viscoelasticity where the study of bulk and surface
waves is a well-developed subject with a wealth of results
obtained over the years. However, this linear framework does not
meet the needs of the actual technological advances in
non-invasive techniques of investigation and medical imaging.
These techniques, based on ultrasound \cite{Sinkus}, are bringing
wave motion to the forefront of imaging and therapy in many areas
of medicine. At the same time, the apparatus used must now rely on
nonlinear constitutive assumptions in order to account for the
pre-loading and large stretches found in living soft tissues; see
the recent review by Hoskins  \cite{Hosk07}. In particular, we
emphasize that pre-loads and pre-deformations are fundamental for
reducing the dynamic range of object stiffness. Indeed,
compression of soft tissues before imaging increases contrast and
reduces de-correlation noise \cite{fatemi1, fatemi2}.

Here we study the propagation of small amplitude waves in certain
iso\-tro\-pic and incompressible nonlinearly viscoelastic solids,
with a view to investigating their stability when subject to large
deformations. The solids under consideration are characterized by
a Cauchy stress tensor $\vec{T}$ depending only on the
Cauchy-Green deformation tensor $\vec{B}$ and on the symmetric
part $\vec{D}$ of the velocity gradient. This class of materials
is usually referred to as \emph{simple materials of differential
type of grade 1}; see Truesdell and Noll \cite{Tru}. This is a
basic class of models in nonlinear viscoelasticity; it accounts
for classical effects like creep and recovery, as in Kelvin-Voigt
linear viscoelasticity, but cannot describe stress relaxation.
This class contains the so-called Mooney-Rivlin viscoelastic
material \cite{Beatty} and the incompressible version of the model
proposed by Landau and Lifschitz in their book on the theory of
elasticity \cite{Landau}.

In Section 2 we summarize the basic governing equations and
constitutive assumptions for these materials. We devote Section 3
to deriving the general form of the incremental equations of
motion in a deformed viscoelastic solid. Then we specialize the
analysis to two-dimensional motions and find conditions for
time-averaged dissipation in time-periodic homogeneous motions. In
Section 4 we study bulk wave propagation and material stability;
we find that the combination of time-averaged dissipation and
strong ellipticity of the static deformation results in material
stability. In Section 5 we consider surfaces waves and geometric
stability; we find some explicit results when we specialize the
analysis to a Mooney-Rivlin solid with Newtonian viscosity.

%A summary of the results is presented along with some concluding remarks in Section 6.

%%%%%%%%%%%%%%%%%%%%%%%%%%

\section{Basic equations}

%%%%%%%%%%%%%%%%%%%%%%%%%

%======================

\subsection{Kinematics}

%======================

Consider a continuous body whose stress-free reference
configuration is denoted $\mathcal{B}_r$ and in which material
points are labelled in terms of their position vectors $\vec{X}$.
In the current (deformed) configuration at time $t$, denoted
$\mathcal{B}$, $\vec{X}$ occupies the position $\vec{x}$, and the
\emph{motion} from $\mathcal{B}_r$ to $\mathcal{B}$ is described
by the bijection mapping $\bm\chi$ such that
\begin{equation}
\vec{x}=\bm\chi(\vec{X},t).\la{e1}
\end{equation}

The \emph{deformation gradient} associated with the motion,
denoted $\vec{F}$, is defined as
\begin{equation}
\vec{F}=\Grad\vec{x},\la{e2}
\end{equation}
where Grad is the gradient operator in $\mathcal{B}_r$, and the
\emph{velocity} $\vec{v}$ of a material particle is defined as
\begin{equation}
\vec{v}= \frac{\partial \vec{x}}{\partial t}\equiv
\frac{\partial\bm\chi}{\partial t}(\vec{X},t).\la{e3}
\end{equation}
It follows that $\partial \vec{F} / \partial t =\vec{L F}$, where
\begin{equation}
\vec{L}=\grad\vec{v},\la{e4}
\end{equation}
with $\vec{v}$ regarded as a function of $\vec{x}$ and $t$, is the
\emph{velocity gradient}. Its symmetric part is the
\emph{strain-rate tensor} $\vec{D}$, given by
\begin{equation}
\vec{D}=\tfrac{1}{2}(\vec{L}+\vec{L}^T),\la{e5}
\end{equation}
where the superscript $^T$ denotes the transpose. Finally, the
left and right Cauchy-Green deformation tensors are defined by
\begin{equation}
\vec{B}=\vec{FF}^T, \qquad \vec{C}=\vec{F}^T\vec{F}, \la{e6}
\end{equation}
respectively,
and we note that
\begin{equation}
\partial \vec{C} / \partial t = 2 \vec{F}^T\vec{DF}.\la{e7}
\end{equation}

For an \emph{incompressible} material only isochoric motions are
permitted, in which case the constraints
\begin{equation}
\det\vec{F} = 1, \qquad \tr\vec{L}=\tr\vec{D}=0,\la{e9}
\end{equation}
are enforced at all times. The latter condition is equivalent to
\begin{equation}
\div\vec{v}=0,\la{e10}
\end{equation}
where div is the divergence operator in $\mathcal{B}$.

For an incompressible material we now define eight independent
invariants of the two tensors $\vec{B}$ and $\vec{D}$ by
\begin{eqnarray}
& & I_1 = \tr\vec{B}, \quad I_2=\tr(\vec{B}^{-1}), \quad
I_5=\tr(\vec{DB}), \quad I_6=\tr(\vec{D B}^2),\notag \\
& & I_7 = \tr(\vec{D}^2), \quad  I_8 = \tr(\vec{D}^2\vec{B}),\quad
I_9=\tr(\vec{D}^2\vec{B}^2), \quad
I_{10}=\tr(\vec{D}^3),\qquad\quad\la{e13}
\end{eqnarray}
noting that the invariants $I_3 = \det\vec{B}=1$ and  $I_4
=\tr\vec{D}=0$ have been omitted from the list by virtue of
\eqref{e6} and \eqref{e9}.

%===================================================

\subsection{Constitutive law and equation of motion}

%===================================================

For an incompressible isotropic material with \emph{Cauchy stress
tensor} $\vec{T}$ depending on $\vec{B}$ and $\vec{D}$ only, the general
representation for the constitutive law is \cite[Chap. 11]{GrAd60}
\begin{multline}
\vec{T}=-p\vec{I}+\alpha_1\vec{B}+\alpha_2\vec{B}^2+\alpha_3 \vec{D}\\
 +  \alpha_4(\vec{D B}+\vec{B D})+\alpha_5(\vec{D B}^2 +
\vec{B}^2\vec{D})+\alpha_6\vec{D}^2\\
 +  \alpha_7(\vec{D}^2\vec{B}+\vec{B}
\vec{D}^2)+\alpha_8(\vec{D}^2\vec{B}^2+\vec{B}^2\vec{D}^2),
 \la{e14}
\end{multline}
where $\vec{I}$ is the identity tensor, $p$ is the Lagrange
multiplier associated with the incompressibility constraint, and
$\alpha_i,\, i\in\{1,2,\dots 8\}$, are material functions that
depend on the eight invariants \eqref{e13}:
\begin{equation}
\alpha_i=\alpha_i(I_1, I_2, I_5, I_6, I_7, I_8, I_9, I_{10}).\la{e15}
\end{equation}

The equation of motion in the absence of body forces is
\begin{equation}
{\rm div}\, \vec{T}=\rho \partial^2 \vec{x} / \partial t^2,\la{e16}
\end{equation}
where $\rho$ is the mass density of the material.
Equivalently, it can be written as
\begin{equation}
{\rm Div}\,\vec{S}=\rho \partial^2 \vec{x} / \partial t^2,\la{e18}
\end{equation}
where Div is the divergence operator in ${\mathcal B}_r$, and
$\vec{S}$ is the \emph{nominal stress tensor}, defined here as
\begin{equation}
\vec{S}=\vec{F}^{-1}\vec{T}.\la{e17}
\end{equation}

%=======================

\subsection{Equilibrium}

%=======================

Suppose now that the material is in equilibrium in a deformed
configuration $\bar{\mathcal{B}}$ so that $\vec{v}=\vec{0}$,
$\vec{D}= \vec{0}$. Let all quantities associated with
$\bar{\mathcal{B}}$ be denoted by an overbar.  Then the
deformation is written
\begin{equation}
\bar{\vec{x}} = \bar{\vec{\chi} }(\vec{X}),\la{e19}
\end{equation}
the associated deformation gradient is $\bar{\vec{F}}$, and the
corresponding left Cauchy-Green tensor is denoted $\bar{\vec{B}}$.
The Cauchy stress is
\begin{equation}
\bar{\vec{T}} = -{\bar p}\vec{I}+{\bar\alpha}_1 \bar{\vec{B}}
 + {\bar\alpha}_2 \bar{\vec{B}}^2,\la{e20}
\end{equation}
where
\begin{equation}
{\bar\alpha}_i=\alpha_i({\bar I}_1,{\bar I}_2,0,\dots, 0),\la{e21}
\end{equation}
and ${\bar I}_1,{\bar I}_2$ are the first two principal invariants
of $\bar{\vec{B}}$. Finally, the equilibrium equation may be
written in either of the equivalent forms
\begin{equation}
\div\bar{\vec{T}} = \vec{0}, \qquad \Div\bar{\vec{S}} =
\vec{0},\la{e23}
\end{equation}
where $\bar{\vec{S}}=\bar{\vec{F}} ^{-1}\bar{\vec{T}}$.

%%%%%%%%%%%%%%%%%%%%%%%%%%%%%%%%%%%%%%%%%%%%%%%%%%%%%%%%%%%%%%

\section{Small motion superimposed on a static finite strain}

%%%%%%%%%%%%%%%%%%%%%%%%%%%%%%%%%%%%%%%%%%%%%%%%%%%%%%%%%%%%%%

%==================================

\subsection{Incremental kinematics}

%==================================

We now superimpose a small amplitude motion on the finite static
deformation in the configuration ${\bar{\mathcal B}}$.  Let
$\vec{\dot{x}}(\vec{X},t)$ denote this incremental motion. We then
change variables from $(\vec{X},t)$ to $(\bar{\vec{x}},t)$ and
introduce the \emph{mechanical displacement} vector
$\vec{u}(\vec{x},t)$ defined by
\begin{equation}
\vec{\dot{x}}(\vec{X},t)=\vec{u}(\mathbf{\bar{\bm\chi}}(\vec{X}),t),\la{e25}
\end{equation}
there being no need to distinguish between $\vec{x}$ and
$\bar{\vec{x}}$ in this linearization. The corresponding increment
in the deformation gradient is
\begin{equation}
\vec{\dot{F}} = \Grad\vec{\dot{x}}
  = \vec{H} \bar{\vec{F}},\la{e26}
\end{equation}
where $\vec{H}=\grad\vec{u}$ is the \emph{displacement gradient}.

Because the basic deformation is static, we have $\vec{v} =
\partial (\bar{\vec{x}} + \vec{\dot{x}})/\partial t
 =  \partial \vec{u} / \partial t$, and hence
\begin{equation}
\partial \vec{\dot{F}} / \partial t
   = \Grad\vec{v} = \vec{L}\bar{\vec{F}},\la{e28}
\end{equation}
where $\vec{L} =\grad\vec{v}$ is the velocity gradient defined in
\eqref{e4}, and we have
\begin{equation}
\vec{L} = \partial \vec{H} / \partial t.\la{e30}
\end{equation}

We compute the (linearized) increments in the relevant kinematical
quantities as
\begin{align}
&  \vec{\dot{B}} = \vec{H}\bar{\vec{B}} + \bar{\vec{B}}\vec{H}^T,
& &  \dot{({\overline{\vec{B}^2}})} = \vec{H}\bar{\vec{B}}^2 +
         \bar{\vec{B}} \vec{H} \bar{\vec{B}} + \bar{\vec{B}}\vec{H}^T\bar{\vec{B}}
       + \bar{\vec{B}}^2 \vec{H}^T,
\notag \\
&  \dot{I}_1 =  2\,{\rm tr}\,(\vec{H}\bar{\vec{B}}), & &
\dot{I}_2 = -2\,{\rm tr}\,(\vec{H}\bar{\vec{B}}^{-1}),
\notag \\
&   \dot{I}_5 = {\rm tr}\,(\vec{D}\bar{\vec{B}}), & & \dot{I}_6 =
{\rm tr}\,(\vec{D}\bar{\vec{B}}^2). \la{e34}
\end{align}
Note that the increments in the invariants $I_7,\dots,I_{10}$ are
zero at first order. In fact, because $\vec{D} = (\vec{L} +
\vec{L}^T)/2$ is infinitesimal by \eqref{e30}, we have
\begin{equation}
 I_7 = I_8 = I_9 = I_{10} = 0
\end{equation}
at first order. It follows that the material parameters $\alpha_i$
in the constitutive equation \eqref{e14} need from now on be
considered as functions of four invariants only, namely $(I_1,
I_2, I_5, I_6)$.  Thus,
\begin{equation}
\alpha_i=\alpha_i(I_1, I_2, I_5, I_6),\quad
i\in\{1,\dots,8\}.\label{alpha}
\end{equation}
%========================================================

\subsection{Incremental stress and incremental equations of motion}

%========================================================

Now we increment the constitutive law \eqref{e14}, retaining only
the first-order terms. We find, using \eqref{e34} and the
increment of \eqref{alpha}, that
\begin{align}
\vec{\dot{T}} =
 &  -\dot{p} \vec{I}+{\bar\alpha}_1(\vec{H} \bar{\vec{B}}
 +\bar{\vec{B}}\vec{H}^T)
   +  {\bar\alpha}_2(\vec{H}\bar{\vec{B}}^2+\bar{\vec{B}}\vec{H}\bar{\vec{B}}
    +\bar{\vec{B}}\vec{H}^T\bar{\vec{B}} + \bar{\vec{B}}^2 \vec{H}^T)
 \notag \\
 &  +  {\bar\alpha}_3\vec{D}+{\bar\alpha}_4(\vec{D} \bar{\vec{B}}
  + \bar{\vec{B}}\vec{D})+ {\bar\alpha}_5 (\vec{D} \bar{\vec{B}}^2
   + \bar{\vec{B}}^2 \vec{D})
\notag \\
 &   +  [2{\bar\alpha}_{11}\tr(\vec{H} \bar{\vec{B}})
    - 2{\bar\alpha}_{12}\tr(\vec{H} \bar{\vec{B}}^{-1})
       + {\bar\alpha}_{15}\tr(\vec{D} \bar{\vec{B}})
         + {\bar\alpha}_{16}\tr(\vec{D}\bar{\vec{B}}^2)] \bar{\vec{B}}
 \notag \\
 & + [2{\bar\alpha}_{21}\tr(\vec{H} \bar{\vec{B}})
  - 2 {\bar\alpha}_{22}\tr(\vec{H} \bar{\vec{B}}^{-1})
    + {\bar\alpha}_{25}\tr(\vec{D} \bar{\vec{B}})
     + {\bar\alpha}_{26}\tr(\vec{D} \bar{\vec{B}}^2)] \bar{\vec{B}}^2,
 \la{e39}
\end{align}
where ${\bar\alpha}_i$ and ${\bar\alpha}_{ij}$ are the values of
$\alpha_{i}$ and $\partial\alpha_i/\partial I_j$, respectively,
evaluated for $\vec{B}=\bar{\vec{B}}$, $\vec{D}=\vec{0}$.

Incrementing the connection \eqref{e17} between Cauchy stress and
nominal stress, and using \eqref{e26}, we obtain the increment in
the nominal stress as
\begin{equation}
\vec{\dot{S}} = \bar{\vec{F}}^{-1}(\vec{\dot{T}} - \vec{H} \bar{\vec{T}}).
\la{e35}
\end{equation}
It follows that the increment in the equation of motion \eqref{e18}, which is
\begin{equation}
\Div\vec{\dot{S}}
 = \rho \partial^2 \vec{\dot{x}}/\partial t^2,\la{e36}
\end{equation}
can equivalently be written as
\begin{equation}
\div(\vec{\dot{T}} - \vec{H}\bar{\vec{T}})
  = \rho \partial^2 \vec{u}/ \partial t^2,\la{e37}
\end{equation}
with $\vec{x}$ and $t$ as the independent variables. This is
coupled with the incremental incompressibility condition
\begin{equation}
\div\vec{u}=0. \la{e37a}
\end{equation}

We recall that the underlying deformation is homogeneous so that
$\bar{\vec{B}}$ is constant, and hence ${\bar\alpha}_i$ and
${\bar\alpha}_{ij}$ are constants, as is $\bar{\vec{T}}$. It
follows that
\begin{equation}
\div(\vec{H}\bar{\vec{T}})=\bar{\vec{T} }(\div
 \vec{H})=\bar{\vec{T} }\grad(\div\vec{u})=\vec{0},
 \la{e40}
\end{equation}
and the equation of motion \rr{e37} reduces to
\begin{equation}
\div\vec{\dot{T}} = \rho \partial^2 \vec{u}/ \partial t^2.
 \la{e41}
\end{equation}
Upon substitution of \rr{e39} into \rr{e41}, we arrive at
\begin{align}
& -\grad\dot{p}
  + \div[{\bar\alpha}_1\bar{\vec{B}} \vec{H}^T
   + {\bar\alpha}_2(\bar{\vec{B}}\vec{H} \bar{\vec{B}}
    + \bar{\vec{B}} \vec{H}^T \bar{\vec{B}}
      +\bar{\vec{B}}^2 \vec{H}^T)]
\nonumber\\
& + \, \div[{\bar\alpha}_3\vec{D}
  + {\bar\alpha}_4(\vec{D}\bar{\vec{B}} + \bar{\vec{B}} \vec{D})
   + {\bar\alpha}_5(\vec{D} \bar{\vec{B}}^2
    + \bar{\vec{B}}^2\vec{D})]
\nonumber\\
&+\, \bar{\vec{B}}\grad[2{\bar\alpha}_{11}\tr(\vec{H}
\bar{\vec{B}})
    - 2{\bar\alpha}_{12}\tr(\vec{H} \bar{\vec{B}}^{-1})
      + {\bar\alpha}_{15}\tr(\vec{D} \bar{\vec{B}})
        + {\bar\alpha}_{16}\tr(\vec{D}\bar{\vec{B}}^2)]
\nonumber\\
&+\, \bar{\vec{B}}^2\grad[2{\bar\alpha}_{21}\tr(\vec{H}
\bar{\vec{B}})
   - 2 {\bar\alpha}_{22}\tr(\vec{H} \bar{\vec{B}}^{-1})
     + {\bar\alpha}_{25}\tr(\vec{D} \bar{\vec{B}})
      + {\bar\alpha}_{26}\tr(\vec{D}\bar{\vec{B}}^2)]
\nonumber \\
& = \rho \partial^2 \vec{u}/ \partial t^2,
      \la{e42}
\end{align}
where we have used the incremental incompressibility condition
\eqref{e37a}.

%========================================================

\subsection{Two-dimensional waves}

%========================================================

Let $\lambda_1^2,\lambda_2^2,\lambda_3^2$ be the eigenvalues of
$\bar{\vec{B}}$ and denote by $x_1,x_2,x_3$ the coordinates
associated with Cartesian axes along the corresponding
eigenvectors. These are the \emph{principal axes} of pre-strain,
and for isotropic materials, as considered here, they are aligned
with the principal axes of the pre-stress.

In the remainder of the paper we focus on two-dimensional waves,
whose spatial variations depend on two principal space variables
only, $x_1$ and $x_2$ say. Hence
\begin{equation}
\vec{u} = \vec{u}(x_1, x_2, t), \qquad
\dot{p} = \dot{p}(x_1, x_2, t),
\end{equation}
and the incremental incompressibility constraints \eqref{e37a} and
\eqref{e10} reduce to
\begin{equation}
u_{1,1} + u_{2,2} = 0, \qquad v_{1,1} + v_{2,2} = 0,
\end{equation}
respectively, where the comma denotes partial differentiation. The
components of $\vec{\dot{T}}$ in the ($x_1, x_2$) plane are
\begin{align}
\dot{T}_{11} = & -\dot{p}
 + 2 (\bar{\alpha}_1 + 2 \lambda_1^2 \bar{\alpha}_2)
   \lambda_1^2 u_{1,1}
 + (\bar{\alpha}_3 + 2 \lambda_1^2 \bar{\alpha}_4 + 2 \lambda_1^4 \bar{\alpha}_5)
     v_{1,1}
\notag \\
& + 2 \lambda_1^2 (\bar{\alpha}_{11} + \lambda_1^2
\bar{\alpha}_{21})
     (\lambda_1^2 u_{1,1} + \lambda_2^2 u_{2,2})
     \notag \\
&
  - 2 \lambda_1^2(\bar{\alpha}_{12} + \lambda_1^2 \bar{\alpha}_{22})
       (\lambda_1^{-2} u_{1,1} + \lambda_2^{-2} u_{2,2})
\notag \\
& + \lambda_1^2 (\bar{\alpha}_{15} + \lambda_1^2
\bar{\alpha}_{25})
     (\lambda_1^2 v_{1,1} + \lambda_2^2 v_{2,2})
     \notag \\
&
 + \lambda_1^2(\bar{\alpha}_{16} + \lambda_1^2 \bar{\alpha}_{26})
       (\lambda_1^{4} v_{1,1} + \lambda_2^{4} v_{2,2}),
\notag \\
\dot{T}_{22} = & -\dot{p}
 + 2 (\bar{\alpha}_1 + 2 \lambda_2^2 \bar{\alpha}_2)
   \lambda_2^2 u_{2,2}
 + (\bar{\alpha}_3 + 2 \lambda_2^2 \bar{\alpha}_4 + 2 \lambda_2^4 \bar{\alpha}_5)
     v_{2,2}
\notag \\
& + 2 \lambda_2^2 (\bar{\alpha}_{11} + \lambda_2^2
\bar{\alpha}_{21})
     (\lambda_1^2 u_{1,1} + \lambda_2^2 u_{2,2})
     \notag \\
&
   - 2 \lambda_2^2(\bar{\alpha}_{12} + \lambda_2^2 \bar{\alpha}_{22})
       (\lambda_1^{-2} u_{1,1} + \lambda_2^{-2} u_{2,2})
\notag \\
& + \lambda_2^2 (\bar{\alpha}_{15} + \lambda_2^2
\bar{\alpha}_{25})
     (\lambda_1^2 v_{1,1} + \lambda_2^2 v_{2,2})
     \notag \\
&
 + \lambda_2^2(\bar{\alpha}_{16} + \lambda_2^2 \bar{\alpha}_{26})
       (\lambda_1^{4} v_{1,1} + \lambda_2^{4} v_{2,2}),
\notag \\
\dot{T}_{12} = &
 \, [\bar{\alpha}_1 + (\lambda_1^2 + \lambda_2^2) \bar{\alpha}_2]
   \lambda_2^2 u_{1,2}
 + [\bar{\alpha}_1 + (\lambda_1^2 + \lambda_2^2) \bar{\alpha}_2]
     u_{2,1}
\notag \\
& + \textstyle{\frac{1}{2}} [\bar{\alpha}_{3} +
       (\lambda_1^2 + \lambda_2^2) \bar{\alpha}_{4} +
       (\lambda_1^4 + \lambda_2^4) \bar{\alpha}_{5}](v_{1,2} + v_{2,1}),
\end{align}
and they do not involve $u_3$.

The incremental equations of motion \eqref{e41} reduce to
\begin{equation}
\dot{T}_{11,1} + \dot{T}_{12,2} = \rho u_{1,tt}, \qquad
\dot{T}_{12,1} + \dot{T}_{22,2} = \rho u_{2,tt}.\label{mtn}
\end{equation}
It is easy to check that these equations decouple from the third
equation of motion $\dot{T}_{13,1} + \dot{T}_{23,2} = \rho
u_{3,tt}$, which involves $u_3$ only.  We therefore take $u_3
= 0$ so that this is satisfied and need not be considered further.
A simple manipulation of \eqref{mtn} then leads to
\begin{equation}
(\dot{T}_{11} - \dot{T}_{22})_{,12}
 + \dot{T}_{12,22} - \dot{T}_{12,11} = \rho (u_{1,2tt} - u_{2,1tt}),\label{mtn2}
\end{equation}
which eliminates $\dot{p}$.

It is now convenient to introduce the material parameters
$\alpha,\gamma,\beta,\delta,\epsilon$, defined by
\begin{align}
\alpha =  & \; [{\bar\alpha}_1 +
{\bar\alpha}_2(\lambda_1^2+\lambda_2^2)]\lambda_1^2,
\notag \\
 \gamma
= & \; [{\bar\alpha}_1 +
{\bar\alpha}_2(\lambda_1^2+\lambda_2^2)]\lambda_2^2,
 \notag \\
2 \beta = & \; {\bar\alpha}_1(\lambda_1^2+\lambda_2^2)
            + {\bar\alpha}_2 (3\lambda_1^4 + 3\lambda_2^4 - 2 \lambda_1^2 \lambda_2^2)
 \nonumber\\
& + 2 {\bar\alpha}_{11}(\lambda_1^2 - \lambda_2^2)^2
  + 2 {\bar\alpha}_{12} \lambda_{1}^{-2} \lambda_{2}^{-2}(\lambda_1^2-\lambda_2^2)^2
  \nonumber\\
& + 2 {\bar\alpha}_{21}(\lambda_1^2 - \lambda_2^2)^2(\lambda_1^2 +
\lambda_2^2)
  + 2 {\bar\alpha}_{22}\lambda_1^{-2} \lambda_2^{-2}
        (\lambda_1^2-\lambda_2^2)^2(\lambda_1^2 + \lambda_2^2),
  \nonumber \\
2 \delta = & \;
{\bar\alpha}_3+{\bar\alpha}_4(\lambda_1^2+\lambda_2^2)
+{\bar\alpha}_5(\lambda_1^4+\lambda_2^4),
  \notag \\
\epsilon = & \; [{\bar\alpha}_{15}
  + ({\bar\alpha}_{16} + {\bar\alpha}_{25})(\lambda_1^2 + \lambda_2^2)
  + {\bar\alpha}_{26} (\lambda_1^2 + \lambda_2^2)^2](\lambda_1^2 - \lambda_2^2)^2.
     \la{e47}
\end{align}
Then, on use of $u_{2,2} = - u_{1,1}$ and $v_{2,2} = - v_{1,1}$,
we obtain
\begin{eqnarray}
\dot{T}_{11} - \dot{T}_{22}
 & =& (\alpha + \gamma + 2 \beta)u_{1,1} + (\epsilon +
 4\delta)v_{1,1},\\
\dot{T}_{12} & =& \alpha u_{2,1} + \gamma u_{1,2} + \delta(v_{1,2}
+ v_{2,1}).
\end{eqnarray}

The incremental incompressibility constraint suggests the
introduction of a scalar potential function $\psi=\psi(x_1,x_2,t)$
such that
\begin{equation}
u_1=\psi_{,2}, \quad u_2=-\psi_{,1}, \qquad v_1=\psi_{,2t}, \quad
v_2=-\psi_{,1t},
 \la{e45}
\end{equation}
use of which enables the equation of motion \eqref{mtn2} to be
cast as an equation for $\psi$, namely
\begin{multline}
 \alpha \psi_{,1111}
  + 2 \beta \psi_{,1122}
  + \gamma \psi_{,2222} \\
   + \delta(\psi_{,1111t} + 2 \psi_{,1122t} + \psi_{,2222t})
    + \epsilon \psi_{,1122t} =
     \rho (\psi_{,11tt} + \psi_{,22tt}).
   \la{e46}
\end{multline}
This is the equation that governs the two-dimensional incremental
motions.

%********************************

\subsection{Dissipation}

%********************************

For a continuum, the work done by external forces is converted
into kinetic energy, stored energy, and dissipated energy.  The
combination of the latter two is measured by the rate of working
of the stresses, which, per unit volume, is
$\tr(\vec{S}\vec{\dot{F}})$. For the incremental motion this can
be written $\tr[(\bar{\vec{S}}+\vec{\dot{S}})\vec{\dot{F}}]$.  The
first term in this sum can be considered as the stored elastic
energy associated with the underlying static deformation, whilst
the second term $\tr(\vec{\dot{S}}\vec{\dot{F}})$ is a measure of
the dissipation associated with the motion (which may include some
additional stored energy).

From \eqref{e35}, \eqref{e28}, the symmetry of $\vec{\dot{T}}$, and
the definition \eqref{e5}, we obtain
\begin{equation}
\tr(\vec{\dot{S}}\vec{\dot{F}})=\tr(\vec{\dot{T}
D})-\tr(\vec{H}\bar{\vec{T}}\vec{L}).\label{dissip}
\end{equation}
For the two-dimensional incremental motions, the two
terms on the right-hand side of \eqref{dissip} may be computed,
respectively, as
\begin{align}
\tr(\vec{\dot{T} D}) =
 & \; \dot{T}_{11} D_{11} + \dot{T}_{22} D_{22} + 2 \dot{T}_{12} D_{12}
 \notag \\
 = & \;
  (\dot{T}_{11} - \dot{T}_{22})v_{1,1} + \dot{T}_{12}(v_{1,2} + v_{2,1})
\notag \\
 = & \;
 (\alpha + \gamma + 2 \beta) u_{1,1} v_{1,1}
  + (\epsilon + 4 \delta)v_{1,1}^2
\notag \\
  & \; + (\alpha u_{2,1} + \gamma u_{1,2})(v_{1,2} + v_{2,1})
    + \delta (v_{1,2} + v_{2,1})^2
\end{align}
and
\begin{multline}
\tr(\vec{H}\bar{\vec{T}}\vec{L}) =
 \bar{T}_{11}(u_{1,1}v_{1,1}+u_{2,1}v_{1,2})+\bar{T}_{22}(u_{1,2}v_{2,1}+u_{2,2}v_{2,2})
   \\
+\bar{T}_{12}(u_{1,2}v_{1,1}+u_{2,2}v_{1,2}+u_{1,1}v_{2,1}+u_{2,1}v_{2,2}).
\end{multline}

In the case of \emph{time-periodic homogeneous motions}, we use
angle brackets to denote the time average over a period; here we
find that
\begin{equation}
\left<\tr(\vec{\dot{T} D})\right> = (\epsilon + 4 \delta)
\left<v_{1,1}^2 \right> + \delta \left<(v_{1,2} +
v_{2,1})^2\right>,\label{TD}
\end{equation}
and that the other terms vanish, as we now show.

First we have
\begin{equation}
 u_{1,1} v_{1,1}  = \psi_{,12} \psi_{,12t} =
 \demi \left(\psi^2_{,12}\right)_{,t},
\end{equation}
whose time average clearly vanishes by periodicity. Next we have
\begin{equation}
 (\alpha u_{2,1} + \gamma u_{1,2})(v_{1,2} + v_{2,1})
  = \demi \alpha  \left(\psi^2_{,11}\right)_{,t}
 +  \demi \gamma  \left(\psi^2_{,22}\right)_{,t}
  - \alpha  \psi_{,11} \psi_{,22t}
   - \gamma \psi_{,22} \psi_{,11t}.
\end{equation}
Here the time averages of the first two terms vanish by
periodicity. To compute the time averages of the last two terms,
we write $\psi$ explicitly. For (two-dimensional) time-harmonic
motions, we may write it in the general form
\begin{equation} \label{psi_hom}
 \psi = C \ee^{\ii \omega (s\, \vec{n \cdot x} - t)}
    +   \overline{C} \ee^{-\ii \omega (\overline{s} \, \vec{n \cdot x} - t)},
\end{equation}
where $C$ is a complex constant, $\omega$ is the real frequency,
$s$ is the complex slowness, $\vec{n}=(n_1,n_2,0)$ is a real unit
vector in the propagation direction, and the overbar denotes the
complex conjugate. Introducing the function $\varphi$ defined by
\begin{equation}
 \varphi = - \omega^2
  \left[C s^2 \ee^{\ii \omega (s \, \vec{n \cdot x} - t)}
    +   \overline{C} \overline{s}^2 \ee^{-\ii \omega (\overline{s} \, \vec{n \cdot
x} - t)}
     \right],
\end{equation}
we obtain the expressions
\begin{equation}
 \psi_{,11} \psi_{,22t} = \psi_{,11} \psi_{,22t} =
   \demi n_1^2 n_2^2 \left( \varphi^2 \right)_{,t},
\end{equation}
which have a zero time average by periodicity.  Similar
calculations show that the time average of
$\tr(\vec{H}\bar{\vec{T}}\vec{L})$ vanishes.

Turning back to the time average of \eqref{TD}, we find, using the
function $\varphi$, that it can be written as
\begin{equation}
\left< \tr(\vec{\dot{T} D})\right> = (\epsilon \, n_1^2 n_2^2 +
\delta) \left< (\varphi_{,t})^2 \right>,\label{<TD>}
\end{equation}
making it clear that the deformed viscoelastic solid is
dissipative under (plane) incremental motions when $\epsilon \,
n_1^2 n_2^2 + \delta > 0$ for all $n_1$, $n_2$ such that $n_1^2 +
n_2^2 = 1$. This is ensured when
\begin{equation} \label{diss}
\delta \geq 0, \qquad \epsilon + 4 \delta \geq 0,
\end{equation}
with at least one of these inequalities being strict.  For the
remainder of the paper, we assume that these inequalities hold.

Having established the conditions for time-averaged dissipation of
time-periodic homogeneous motions, we now investigate stability
issues for a deformed viscoelastic solid occupying, first, the
whole space and, second, a semi-infinite space.

%%%%%%%%%%%%%%%%%%%%%%%%%%%%%%

\section{Material stability}

%%%%%%%%%%%%%%%%%%%%%%%%%%%%%%

First we look at the situation where the perturbation has \emph{no
time dependence}, that is when $\partial / \partial t = 0$. For
all intents and purposes, the viscous effects are not felt then,
and the solid behaves as a purely elastic solid. The corresponding
incremental equation of elastostatics is the specialization of
\eqref{e46} to
\begin{equation}
 \alpha \psi_{,1111}
  + 2 \beta \psi_{,1122}
  + \gamma \psi_{,2222} = 0.
\end{equation}
It is known \cite{dowa90} that this equation is \emph{strongly elliptic}
when
\begin{equation}
 \alpha > 0, \qquad \gamma>0, \qquad
  \beta + \sqrt{\alpha \gamma} > 0,\label{SE}
\end{equation}
and we assume henceforth that these inequalities hold. This
guarantees material stability in the strong ellipticity sense with
respect to incremental static deformations.

Next we study \emph{bulk homogeneous plane waves} because they
provide a natural tool for addressing the question of the material
(bulk) stability of the deformed viscoelastic solid. We therefore
seek solutions of the form
\begin{equation} \label{psi_homog}
 \psi = \psi_0 \ee^{\ii \omega (s \, \vec{n \cdot x} - t)},
\end{equation}
where $\psi_0$ is a constant, $\omega = \omega^+ + \ii \omega^-$
is the \emph{complex frequency}, $s = s^+ + \ii s^-$ is the
\emph{complex scalar slowness}, and $\vec{n}$ is a real
two-dimensional unit vector in the direction of propagation. Note
that this motion is not necessarily time-periodic because
$\omega^-$ may be different from zero. Combining this motion with
the expressions in \eqref{e45}, we see that the displacement,
velocity and stress fields have the same exponential dependence as
$\psi$. The argument of the exponential may be decomposed as
\begin{multline}
 \ii \omega  (s \, \vec{n \cdot x} - t) =
  - \left[ (\omega^+ s^- + \omega^- s^+) \, \vec{n \cdot x} - \omega^- t \right]
 \notag \\
  + \ii \left[ (\omega^+ s^+ - \omega^- s^-) \, \vec{n \cdot x} - \omega^+ t \right].
\end{multline}
The first bracketed term gives the amplitude variations of the
fields, and the second one their phase.

Material stability is ensured when there is \emph{no amplitude
growth for a given phase} \cite{BoHa93}. In other words, when
$(\omega^+ s^- + \omega^- s^+) \, \vec{n \cdot x} - \omega^- t \ge
0$ with $ (\omega^+ s^+ - \omega^- s^-) \, \vec{n \cdot x} -
\omega^+ t =0$. This gives
\begin{equation}
\dfrac{\omega^+ s^- + \omega^- s^+}{\omega^+ s^+ - \omega^- s^-} \omega^+
  - \omega^- \ge 0,
\end{equation}
or equivalently, after removing the positive factor
$[(\omega^+)^2 + (\omega^-)^2]$, and taking the inverse,
\begin{equation} \label{stab}
 \dfrac{s^+}{s^-} \omega^+ - \omega^- \ge 0.
\end{equation}
We now examine the implications of this inequality for a deformed
viscoelastic solid.

Substitute the expression \eqref{psi_homog} for $\psi$ into the
equation of motion \eqref{mtn2}, and separate the real and
imaginary parts to obtain
\begin{align}
& \alpha n_1^4 + 2 \beta n_1^2 n_2^2 + \gamma n_2^4
 + \omega^- (\delta + \epsilon n_1^2 n_2^2) =
  \rho [(v^+)^2 - (v^-)^2], \notag
  \\[0.1cm]
& \omega^+ (\delta + \epsilon n_1^2 n_2^2) = - 2 \rho v^+
v^-,\label{mtn_split}
\end{align}
where $v^\pm$ are real quantities defined by
$ v^+ + \ii v^- = (s^+ + \ii s^-)^{-1}$,
that is
\begin{equation} \label{relazionivs}
v^+ = \frac{s^+}{(s^+)^2 + (s^-)^2}, \qquad
v^- = -\frac{s^-}{(s^+)^2 + (s^-)^2}.
\end{equation}

From equation \eqref{mtn_split}$_2$ we find
\begin{equation}
 \dfrac{v^-}{v^+} = - \dfrac{\omega^+(\delta + \epsilon n_1^2 n_2^2)}
  {2 \rho (v^+)^2}.
\end{equation}
Then, dividing equation \eqref{mtn_split}$_1$ through by
$(v^+)^2$, and using this latter identity, we find an expression
for $\omega^-$. We can also use the identity above to find $s^+ /
s^- = - v^+ / v^-$. We end up with
\begin{align}
& \omega^- = \dfrac{\rho (v^+)^2 - \alpha n_1^4 - 2 \beta n_1^2
n_2^2 - \gamma n_2^4}
 {\delta + \epsilon n_1^2 n_2^2}
  - \dfrac{1}{4} \rho (v^+)^2 (\omega^+)^2 (\delta + \epsilon n_1^2 n_2^2),
  \notag \\
& \omega^+ \dfrac{s^+}{s^-} = 2 \dfrac{\rho (v^+)^2}{\delta +
\epsilon n_1^2 n_2^2}.
\end{align}
Hence the stability condition \eqref{stab} reads
\begin{equation}
4 \dfrac{\rho (v^+)^2 + \alpha n_1^4 + 2 \beta n_1^2 n_2^2 + \gamma n_2^4}
 {\delta + \epsilon n_1^2 n_2^2} +
  \rho (v^+)^2 (\omega^+)^2 (\delta + \epsilon n_1^2 n_2^2) \ge 0.
\end{equation}
This condition is clearly satisfied when both \eqref{diss} and
\eqref{SE} hold.
In other words, \emph{time-averaged dissipation with respect to time-periodic motions,
coupled to strong ellipticity with respect to static deformations,
results in material stability}.

Before we go on to investigate geometric stability,
we pause to consider a classic sub-case of the general
bulk wave \eqref{psi_homog}, namely
the \textit{damped travelling wave} solution.
It is of the form
\begin{equation} \label{damped}
\psi = \psi_0 \ee^{- a t} \cos k(\vec{n \cdot x} - c t),
\end{equation}
where $a$ is the damping factor, $k$ is the wavenumber,
and $c$ is the speed.
It is an important subclass of \eqref{psi_homog}, obtained by
taking
\begin{equation} \label{damped_condition}
\omega^+ s^- + \omega^- s^+ = 0,
\end{equation}
so that there is no spatial attenuation of the amplitude. Then
\eqref{psi_homog} specializes to \eqref{damped} by making the
identifications
\begin{equation}
a = - \omega^-, \qquad k= - a / v^-, \qquad c = v^+.
\end{equation}
Also, \eqref{damped_condition} gives
$\omega^+ = (s^+/s^-)a = -(v^+/v^-)a$, so that the dispersion equations
\eqref{mtn_split} reduce to
\begin{align} \label{damped_split}
& \rho c^2 = \alpha n_1^4 + 2 \beta n_1^2 n_2^2 + \gamma n_2^4
 - (\delta + \epsilon n_1^2 n_2^2)^2 k^2 / (4 \rho),
  \notag
  \\[0.1cm]
& 2 \rho a = (\delta + \epsilon n_1^2 n_2^2)k^2.
\end{align}

If damped travelling waves \eqref{damped} can be generated in a
viscoelastic solid, then these equations provide a means to
determine the constitutive parameters by variation of the
propagation direction and of the underlying deformation. In
particular, if the response of the solid shows a dependence of the
damping factor $a$ on the direction of propagation, then the
constitutive model must be such that $\epsilon \ne 0$, according
to \eqref{damped_split}$_2$. By \eqref{e47}$_5$, this means that
the constitutive parameters $\alpha_1$ and $\alpha_2$ cannot both
be completely independent of the invariants $I_5$ and $I_6$.
Therefore, only  certain (quite complex) constitutive models can
display an influence of the propagation direction on the damping
factor. If the model is such that $\epsilon>0$, then the
directions of maximal dissipation are along the bisectors of the
principal directions, and those of minimal damping are aligned
with the principal axes (and vice-versa if the model is such that
$\epsilon<0$).

Conversely,  if the response of the solid shows that the damping
factor $a$ is independent of the direction of propagation for
damped travelling waves, then the constitutive parameters
$\alpha_1$ and $\alpha_2$ can both be completely independent of
the invariants $I_5$ and $I_6$.

%%%%%%%%%%%%%%%%%%%%%%%%%%%%%%

\section{Geometric stability}

%%%%%%%%%%%%%%%%%%%%%%%%%%%%%%

To study surface stability, we consider inhomogeneous motions in
the half-space $x_2 > 0$ with boundary $x_2=0$ in the
$(x_1,x_2)$-plane, the deformation corresponding to pure
homogeneous strain with the principal axes of strain coincident
with the Cartesian axes. On the surface $x_2=0$ we assume that the
incremental surface tractions vanish, i.e.
\begin{equation}
({\bf\dot{\vec{T}}}-\vec{H}\bar{\vec{T}})_{21}=({\bf\dot{\vec{T}}}
 - \vec{H}\bar{\vec{T}})_{22}=0.\la{e52}
\end{equation}
The shear traction condition leads, after some manipulation, to a
condition involving $\psi$, namely
\begin{equation}
\gamma(\psi_{,22}-\psi_{,11})+\sigma_2\psi_{,11}
+\delta(\psi_{,22t}-\psi_{,11t})=0\la{e53}
\end{equation}
on $x_2=0$, where $\sigma_2$ is the (uniform) principal stress
normal to the boundary in basic state of deformation, i.e.
$\sigma_2 \equiv \bar{T}_{22}$.

In order to express the normal component of the incremental
traction on the boundary in terms of $\psi$ it is first necessary
to differentiate the latter equation in \rr{e52} along the
boundary, i.e. with respect to $x_1$, and then make use of the
first component of the equation of motion to eliminate
$\partial\dot{p}/\partial x_1$ (assuming this holds on the
boundary). After further manipulations this leads to
\begin{equation}
(2\beta+\gamma-\sigma_2)\psi_{,112}+\gamma\psi_{,222}
+(\epsilon+3\delta)\psi_{,112t}+\delta\psi_{,222t}
-\rho\psi_{,2tt}=0\la{e54}
\end{equation}
on $x_2=0$. When the viscous terms are absent
($\delta=\epsilon=0$) the boundary conditions are then precisely
those given by Dowaikh and Ogden \cite{dowa90} for the purely
elastic case.

We now consider waves of the inhomogeneous form
\begin{equation}
\psi=\psi_0\text{e}^{\text{i}(kx_1-\omega
t)}\text{e}^{-ksx_2},\la{psi-surface}
\end{equation}
where $k,\omega$ and $s$ may be complex.  However, we impose the
following \emph{propagation inequalities}
\begin{equation}
\text{Re}(k) \geq 0,\qquad \text{Im}(k)\geq 0,\qquad
\text{Re}(\omega) \geq 0, \la{criteria1}
\end{equation}
so that the wave propagates in the \emph{positive} $x_1$ direction
at the interface $x_2 = 0$ and attenuates (if at all) in the
positive $x_1$ direction. Additionally, we set
\begin{equation}
\text{Re}(ks) > 0, \la{criteria2}
\end{equation}
so that the wave decays away from the boundary $x_2=0$ (the
\emph{localization condition}). Finally, we pay special attention
to the sign of $\text{Im}(\omega)$; clearly, if
\begin{equation}
 \text{Im}(\omega) < 0,\la{criteria3}
\end{equation}
then the wave is damped (decays in time); if $\text{Im}(\omega)>0$
it blows up in time, indicating the onset on instability, at least
in the linearized theory. We refer to \eqref{criteria3} as the
\emph{stability condition}.  If $\text{Im}(\omega)=0$ there is
neither growth nor decay in time.

On substitution of \eqref{psi-surface} into equation \eqref{e46}
we obtain a bi-quadratic for $s$, which can be written compactly
in the form
\begin{equation}
\hat{\gamma}s^4-(2\hat{\beta}-\hat{\Omega}) s^2+\hat{\alpha}-
\hat{\Omega} =0,\la{s-quartic}
\end{equation}
where we have introduced the notations
\begin{equation}
\hat{\alpha}=\alpha-\text{i}\omega\delta,\quad
2\hat{\beta}=2\beta-2\text{i}\omega\delta-\text{i}\omega\epsilon,\quad
\hat{\gamma}=\gamma-\text{i}\omega\delta,\quad
\hat{\Omega} = \rho\omega^2/k^2.\la{alphabetagammaOmega}
\end{equation}

The general solution of the equation of motion may be written in
the form
\begin{equation}
\psi=\text{e}^{\text{i}(kx_1-\omega
t)}(A \text{e}^{-ks_1x_2} + B\text{e}^{-ks_2x_2}),\la{psi-gen-sol}
\end{equation}
where $A$ and $B$ are constants and $s_1$ and $s_2$ are the
solutions of \eqref{s-quartic} that satisfy \eqref{criteria1},
\eqref{criteria2},  and \eqref{criteria3}. Substitution of
\eqref{psi-gen-sol} into the boundary conditions \eqref{e53} and
\eqref{e54} then gives the two equations
\begin{align}
& [\hat{\gamma}(s_1^2+1) - \sigma_2]A
  + [\hat{\gamma}(s_2^2+1)-\sigma_2]B = 0,
  \notag \\[0.1cm]
& [2\hat{\beta}+\hat{\gamma}-\sigma_2-
\hat{\Omega}-\hat{\gamma}s_1^2]s_1 A
 + [2\hat{\beta}+\hat{\gamma}-\sigma_2-\hat{\Omega}-\hat{\gamma}s_2^2]s_2 B=0,
\end{align}
for $A$ and $B$.

After removal of the factor $s_1-s_2$ the determinant of
coefficients yields the \emph{dispersion equation}, which, on use of the sum
and product of the roots of \eqref{s-quartic}, reads
\begin{equation}
(\hat{\gamma}-\sigma_2)^2-\hat{\gamma}(\hat{\alpha}-\hat{\Omega})
-\hat{\gamma}s_1s_2(2\hat{\beta}+2\hat{\gamma}-2\sigma_2-\hat{\Omega})=0.\la{dispersion}
\end{equation}
The product $s_1s_2$ has not been replaced since there are two
possible solutions of
$s_1^2s_2^2=(\hat{\alpha}-\Omega)/\hat{\gamma}$ and this needs
careful evaluation.  This dispersion equation reduces to the
elasticity result obtained by Dowaikh and Ogden \cite{dowa90} on
setting $\delta=\epsilon=0$ and $\omega$ and $k$ real.  We remark
here that if the case $s_1=s_2$ is considered separately and the
solution \eqref{psi-gen-sol} amended accordingly then the
associated dispersion condition reduces to $\sigma_2=\pm
2\hat{\gamma}$.  It can be shown that this also follows from the
appropriate specialization of \eqref{dispersion}.  However, since
$\sigma_2$ is real and, in general, $\hat{\gamma}$ is complex,
this cannot be satisfied unless $\text{Re}(\omega)=0$ or
$\delta=0$. In the purely elastic case the corresponding special
solution is $\sigma_2=\pm 2\gamma$ \cite{dowa90}.

From \eqref{alphabetagammaOmega} it follows that
\begin{equation}
2\hat{\beta}-\hat{\alpha}-\hat{\gamma}=2\beta-\alpha-\gamma-\text{i}\epsilon\omega.
\end{equation}
Now, in the context of elasticity, materials for which
$2\beta-\alpha-\gamma=0$ form a special class and lead to
simplifications in the analysis.  Similar simplifications occur
here if we focus on viscoelastic solids for which
\begin{equation}
2\beta = \alpha + \gamma, \qquad \epsilon=0, \label{specialmodel}
\end{equation}
and we assume henceforth that the material model is specialized in
accordance with \eqref{specialmodel}. Then, \eqref{s-quartic}
factorizes to give
\begin{equation}
(s^2-1)[\hat{\gamma}s^2-(\hat{\alpha}-\hat{\Omega})]=0.
\end{equation}
One root consistent with the restrictions \eqref{criteria1},
\eqref{criteria2} is $s = 1$ and we refer to this as $s_1$.
%
%The second root requires a little manipulation.

%If we write $\Omega=\Omega^+ +\text{i}\Omega^-$ and similarly for
%$\hat{\alpha}$ and $\hat{\gamma}$ then $|s|$ is given by
%\begin{equation}
%|s|^4=\frac{(\hat{\alpha}^+-\Omega^+)^2+(\hat{\alpha}^--\Omega^-)^2}
%{(\hat{\gamma}^+)^2+(\hat{\gamma}^-)^2},
%\end{equation}
%while the argument of $s$ is $\theta+\pi/2$, where
%$\theta\in(0,\pi/2)$ satisfies
%\begin{equation}
%\tan 2\theta=\frac{(\hat{\alpha}^--\Omega^-)\hat{\gamma}^+-(\hat{\alpha}^+-\Omega^+)\hat{\gamma}^-}
%{(\hat{\alpha}^+-\Omega^+)\hat{\gamma}^++(\hat{\alpha}^--\Omega^-)\hat{\gamma}^-},
%\end{equation}
%so that
%\begin{equation}
%s=|s|(-\sin\theta+\text{i}\cos\theta).
%\end{equation}
%Let this be designated $s_2$.  To satisfy \eqref{criteria2} we
%require $k^+\cos\theta&gt;k^-\sin\theta$.
%
There are two possibilities for the second root,
\begin{equation} \la{s2}
 s_2 = \pm\sqrt{\dfrac{\hat{\alpha} - \hat{\Omega}}{\hat{\gamma}}}.
\end{equation}
A test must be conducted by computing $k s_2$ for each possibility
and checking whether the localization requirement
\eqref{criteria2} is satisfied. Here the square root symbol
designates the complex number with square equal to $(\hat{\alpha}
- \hat{\Omega})/\hat{\gamma}$ and positive real part. In any case,
the dispersion equation \eqref{dispersion} can be re-cast as a
cubic in $s_2$, namely
\begin{equation}
 s_2^3 + s_2^2 + (3-2 \hat{\sigma}_2)s_2 - (1-\hat{\sigma}_2)^2 = 0,
\label{disp_neo_sigma}
\end{equation}
where $\hat{\sigma}_2 = \sigma_2 / \hat{\gamma}$. Note that this
cubic is not obtained by a squaring process, in contrast to the
cubic obtained by Currie et al. \cite{curr77}. It does not contain
spurious roots \emph{a priori} and it is therefore legitimate to
check the validity of each of its three roots against conditions
\eqref{criteria1}, \eqref{criteria2}, and \eqref{criteria3} once
$k$ (or $\omega$) has been deduced from \eqref{s2} for a given
$\omega$ (or $k$).

However, the behaviour of the roots is highly dependent on the
material parameters and on the pre-stress and pre-strain, and
little can be concluded in general. In order to make progress and
provide an illustrative example, we first specialize the
analysis further to a \emph{Mooney-Rivlin solid with Newtonian
viscosity}, with constitutive equation
\begin{equation} \label{neovis}
 \vec{T} = -p \vec{I} + (C_1 + C_2 I_1) \vec{B}
  - C_2 \vec{B}^2 + \nu \vec{D},
\end{equation}
where $C_1$, $C_2$, and $\nu$ are positive constants.
Then the parameters $\alpha$, $\gamma$, $\beta$, $\delta$, $\epsilon$
of \eqref{e47} reduce to
\begin{align}
& \alpha = (C_1 + C_2 \lambda_3^2) \lambda_1^2, \quad \gamma =
(C_1 + C_2 \lambda_3^2) \lambda_2^2, \quad
2\beta = (C_1 + C_2 \lambda_3^2) (\lambda_1^2 + \lambda_2^2), \notag \\
& \delta = 0, \quad \epsilon = 0,
\end{align}
making it clear that this solid belongs to the class \eqref{specialmodel}.
The quantity $\mu \equiv C_1 + C_2$ is its \emph{static shear modulus}
and $\nu$ is its \emph{dynamic viscosity}.
Next, we specialize the pre-deformation and pre-stress
to a \emph{plane strain with no normal load},
\begin{equation}
 \lambda_1 = \lambda, \qquad \lambda_2 = \lambda^{-1}, \qquad
  \lambda_3 = 1, \qquad \sigma_2 = 0,
\end{equation}
where $\lambda$ is the stretch ratio in the $x_1$ direction. By
taking $\nu = 0$, this set-up allows for direct comparison with
the purely elastic Mooney-Rivlin case, for which Biot \cite{Biot63}
showed that the critical compressive stretch for surface
instability is $\lambda_\text{cr} = 0.5437$. Also, $s_2$ is now a
root of the cubic $s_2^3 + s_2^2 + 3 s_2 - 1 = 0$, independent of
the material parameters and the pre-deformation. Explicitly, $s_2$
is among the three solutions of this cubic, which are
\begin{equation}
s_2^0 = 0.2956, \qquad s_2^\pm = -0.6478 \pm 1.721 \,
\ii,\label{roots}
\end{equation}
and the dispersion equation is deduced from \eqref{s2} as
\begin{equation}
 \label{disp_neo}
 s_2 = \pm \sqrt{\dfrac{\lambda^2 - \ii \omega \nu/\mu - \rho \omega^2/(\mu k^2)}
         {\lambda^{-2} - \ii \omega \nu / \mu}}.
\end{equation}

Despite the strong simplifying assumptions made here, the
possibilities for solutions to the surface stability problem
remain rich and varied because of the possible complex nature of
the wave number and of the frequency.

%===============================================

\subsection{Real frequency, complex wave number}

%===============================================

\begin{figure}
\centering \epsfig{figure=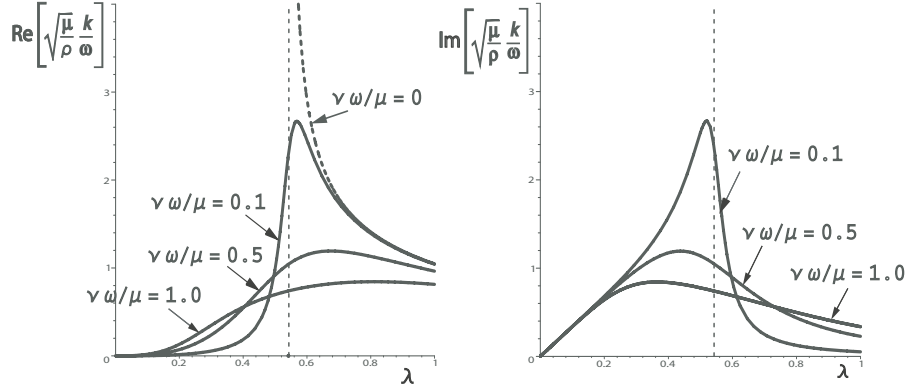, height=.45\textwidth,
width=\textwidth} \caption{Dimensionless slowness and damping
factor for a surface wave with real frequency in a Mooney-Rivlin
viscoelastic solid subject to plane strain. The analysis shows
that the half-space becomes unstable when compressed by more than
about 46\% (vertical asymptote) and this confines the validity of
the curves to the right of the vertical dashed line.}
\label{fig_1}
\end{figure}

First we take $\omega$ real ($\omega > 0$). Then there is neither
growth nor decay in time. In other words, taking $\omega$ real is
not appropriate for the study of stability. Nevertheless, we may
investigate the possibility of a \emph{surface wave} existing,
i.e. a solution in the form \eqref{psi-surface} satisfying all
four conditions \eqref{criteria1}, \eqref{criteria2}. When we
choose $s_2 = s_2^0 = 0.2956$ as the root from \eqref{roots}, we
find that these conditions are equivalent to
\begin{equation}
 \label{cond1}
 \text{Re} \left( \sqrt{\dfrac{\mu}{\rho}} \dfrac{k}{\omega} \right) > 0, \qquad
 \text{Im} \left( \sqrt{\dfrac{\mu}{\rho}} \dfrac{k}{\omega} \right) > 0.
\end{equation}
Figure \ref{fig_1} shows the variations of these quantities with
respect to $\lambda$ for several values of the dimensionless
parameter $\nu \omega / \mu$. In the first figure, the dashed
curve represents the (non-dimensional) slowness in the purely
elastic case ($\nu = 0$), with a vertical asymptote at the
critical stretch $\lambda_\text{cr} = 0.5436$ where the speed
drops to zero. The introduction of viscosity removes this
singularity and a surface wave may propagate for the whole
compressive range, unless of course the half-space becomes
unstable (see below). In the second figure there is no curve at
$\nu = 0$ because the purely elastic wave is not damped
\cite{Flav63}.

When we take either $s_2 = s_2^\pm$ as the root from
\eqref{roots}, we find that the conditions \eqref{criteria1},
\eqref{criteria2} cannot be satisfied simultaneously. Hence, there
is only one possibility for a surface wave to propagate over a
deformed viscoelastic Mooney-Rivlin solid, that which tends to the
Rayleigh surface wave solution when the viscosity tends to zero.
This is in accord with the results of Romeo \cite{rome02} in
linear elasticity (no finite pre-deformation).

%=================================================

\subsection{Complex frequency, complex wave number}

%=================================================

When we allow both the frequency and the wave number to be
complex, we find that the imaginary part of $\omega$ is negative
only in the range where $\lambda^2 - \lambda^{-2} s_2^2 \geq 0$,
for $s_2 = s_2^0 = 0.2956$; the other two roots $s_2^\pm$ of
\eqref{roots} do not yield any conclusion with respect to
stability analysis. When $\lambda^2 - \lambda^{-2} s_2^2 < 0$,
i.e. when $\lambda < \lambda_\text{cr} = 0.5436$, the imaginary
part of $\omega$ is positive, indicating instability. When
$\lambda = \lambda_\text{cr} = 0.5436$, both the real and
imaginary parts of $\omega$ are zero, as can be checked from the
dispersion equation \eqref{disp_neo}. The conclusion is then that
the half-space becomes unstable when the complex speed $\omega /
k$ is zero, just as in the purely elastic case. This is in accord
with the correspondence principle of Biot \cite{Biot65}.

\end{document}